\documentclass[aps, prd, twocolumn,print, showpacs, nofootinbib]{revtex4}
\usepackage{amsmath}
\usepackage{cases}
\usepackage{txfonts}
\usepackage{amssymb}
\usepackage{color}
\usepackage{setspace}%
\usepackage{mathrsfs}
\usepackage{graphicx}
\usepackage{epsfig}
\usepackage{dcolumn}
\usepackage{bm}
\begin{document}
\begin{spacing}{1.0}
\title{Net reaction rate and neutrino emissivity for the Urca process in departure from chemical equilibrium}

\author{Wei-Hua Wang,$^{1,}$\footnote{wangweihua@mails.ccnu.edu.cn} Xi Huang,$^{1,2,3}$ and Xiao-Ping Zheng$^{1,}$\footnote{zhxp@phy.ccnu.edu.cn}}

\address{
$^{1}$Institute of Astrophysics, Central China Normal University, Wuhan 430079, China\\
$^{2}$Key Laboratory of Quark and Lepton Physics (Ministry of
Education), Central China Normal University, Wuhan 430079, China\\
$^{3}$School of Electronic and Electrical Engineering, Wuhan Textile
University, Wuhan 430073, China\\}


\begin{abstract}
We discuss the effect of compression on Urca shells in the ocean and
crust of accreting neutron stars, especially in superbursting
sources. We find that Urca shells may be deviated from chemical
equilibrium in neutron stars which accrete at several tenths of the
local Eddington accretion rate. The deviation depends on the energy
threshold of the parent and daughter nuclei, the transition
strength, the temperature, and the local accretion rate. In a
typical crust model of accreting neutron stars, the chemical
departures range from a few tenths of $k_{B}T$ to tens of $k_{B}T$
for various Urca pairs. If the Urca shell can exist in crusts of
accreting neutron stars, compression may enhance the net neutrino
cooling rate by a factor of about $1\sim2$ relative to the neutrino
emissivity in chemical equilibrium. For some cases, such as Urca
pairs with small energy thresholds and/or weak transition strength,
the large chemical departure may result in net heating rather than
cooling, although the released heat can be small. Strong Urca pairs
in the deep crust are hard to be deviated even in neutron stars
accreting at the local Eddington accretion rate.

\end{abstract}

\maketitle

\section{INTRODUCTION} \label{S:intro}
Superbursts are rare, long lasting and energetic x-ray flares which
originate from accreting neutron stars in low-mass x-ray binaries
(LMXBs)~\cite{Cornelisse:2000,Zand:2004,Strohmayer:2002}. The flares
rise in a few seconds and decay in a few hours in an
exponential-like way, and the fluences can be as large as
$10^{42}~\rm{erg}$. These characteristics distinguish them from
``ordinary'' Type-\uppercase\expandafter{\romannumeral1} x-ray
bursts, which are $1000$ times shorter, less energetic, and more
frequent. Besides, the local accretion rates of the superbursting
sources are supposed to be
$\dot{m}=(0.1-0.3)\dot{m}_{Edd}$~\cite{Wijnands:2001}, where
$\dot{m}_{Edd}\simeq 10^{5}~\rm{g~cm^{-2}~s^{-1}}$ is the local
Eddington accretion rate.

The current superburst model proposes that continual accretion
compresses the preexisting material deeper into the crust, raises
the chemical potential of electrons, $\mu_{e}$, and induces
nonequilibrium reactions if the element transformation is
energetically favorable. An energy of heat of about $\approx
1-2~\rm{MeV}$ per accreted nucleon will be deposited through
electron capture (EC), neutron emissions, and pycnonuclear
fusion~\cite{Haensel:1990,Haensel:2003,Gupta:2007,Haensel:2008}.
Part of this energy flows upward to heat the neutron star ocean,
where the superbursts are supposed to be triggered by unstable
burning of
$^{12}\rm{C}$~\cite{Woosley:1976,Taam:1978,Brown:1998,Cumming:2001,Strohmayer:2002,Cooper:2009}
in column depth $y_{ign}\approx
(0.5-3)\times10^{12}~\rm{g/cm^{-2}}$~\cite{Cumming:2006}.

The high temperature dependence of carbon
burning~\cite{Caughlan:1988} makes the ignition sensitive to deep
crustal heating; accordingly, superburst observations have the
potential to constrain the combined study of crustal heating and
neutrino cooling of the crust and core~\cite{Keek:2012}, and also to
act as probes of nuclear physics~\cite{Cumming:2006}. However, the
actual crust temperature is set by deep crustal heating, neutrino
cooling in the crust and core, and heat transport from the interior.
Although previous works showed that the thermal properties of the
neutron star core (for example, core neutrino emissivity) are more
important~\cite{Brown:2004,Cooper:2005} for ignition conditions, a
study on crust neutrino cooling may also be meaningful because crust
neutrino cooling appears in a shallower column depth than core
neutrino cooling.

Recently, Schatz $et~al.$ proposed that the Urca shell may exist in
accreting neutron star crusts~\cite{Schatz:2014}. The Urca cooling
mechanism~\cite{Gamow:1941}, which proceeds back and forth between
specific nuclei (or nucleons) and takes a great amount of energy
away via neutrinos (and antineutrinos)~\cite{Deibel:2016}, has been
considered by Tsuruta and Cameron (TC70) in white dwarfs
\cite{Tsuruta:1970} and in
Type-\uppercase\expandafter{\romannumeral1}a
supernovae~\cite{Paczynski:1972,Woosley:1986}. In TC70, the authors
proposed that the thermal rounding of the Fermi surface or a
vibrational oscillation will make phase space available for the Urca
process to produce the Urca shell. For neutron stars, the thermal
energy is pretty small compared with the Fermi energy, thus the
standard neutron star model uses the zero-temperature approximation,
in which case the parent nucleus $(Z,A)$ are transformed by capture
of degenerate electrons into the daughter nucleus $(Z-1,A)$, while
the daughter nucleus $(Z-1,A)$ cannot decay through
$(Z-1,A)\rightarrow (Z,A)+e^{-}+\overline{\nu}_{e}$ because no phase
space is available to re-emit the captured electrons. By considering
the relatively high temperature ($T>10^{8}~\rm{K}$) and possible
low-lying excited states $E_{x}\lesssim k_{B}T$~\cite{Schatz:2014},
Schatz $et~al.$ showed that the crust Urca neutrino emissivity may
be comparable with the crustal heating rate at the temperature in
accreting neutron crusts (see Fig. 3 in their paper); thus the Urca
pairs may have the potential to cool the outer crust, and make the
surface layers thermally decouple from the deeper crust. If this is
true, it will be a great challenge to current thermonuclear bursts
models~\cite{Keek:2011,Woosley:2004}.

Deibel $et~al.$ continued this work. They identified 85 odd-\emph{A}
isotopes that form Urca pairs in the neutron star ocean by combining
with the crust Urca pairs identified by Schatz $et~al.$ and the Urca
pairs abundances in x-ray burst and superburst ashes. They concluded
that ocean Urca pairs will not have an impact on carbon ignition,
while the strong crust Urca pairs may lower the ocean's steady-state
temperature and increase carbon ignition depths \cite{Deibel:2016}.

We propose that, at such a large local mass accretion rate
$\dot{m}=(0.1-0.3)\dot{m}_{Edd}$ in superbursting sources, the
effect of compression on the Urca shell should be considered.
Although it has been reported that superbursts may occur in neutron
stars at near-local-Eddington mass accretion rate~\cite{Zand:2004},
a local accretion rate $\dot{m}=(0.1-0.3)\dot{m}_{Edd}$ is large
enough to compress the Urca shell on a timescale shorter than the
weak-interaction timescale. Thus, there may be a departures from
chemical equilibrium for the the Urca processes. We strongly urge a
study on deviated Urca processes. The reaction rate and neutrino
emissivity should be calculated.

In this work, we present the ``deviated shell" to distinguish it
from the usual Urca shell in chemical equilibrium. In Sec.
\uppercase\expandafter{\romannumeral2}, we present phase-space
integrals of the reaction rate and neutrino emissivity. In Sec.
\uppercase\expandafter{\romannumeral3}, we present the effect of
accretion-driven compression on the Urca shell. Results and some
discussions are presented in Sec.
\uppercase\expandafter{\romannumeral4}.

\section{Phase space integrals}
We consider the EC and $\beta^-$ decay cycle between certain pairs
of nuclei $(Z,A)$ and $(Z-1,A)$. $\emph{Z}$ is the charge number and
$\emph{A}$ is the mass number.

$(Z,A)+e^-\rightarrow(Z-1,A)+\nu_{e}$,

$(Z-1,A)\rightarrow(Z,A)+e^-+\overline{\nu}_{e}$.

We assume the transition is only from ground state to ground state
or from state $i$ of parent nuclei to state $f$ of daughter nuclei,
$E_{i}$ and $E_{f}$ are the energies for state $i$ of nucleus
$(Z,A)$ and state $f$ of nucleus $(Z-1,A)$ , where $E_{i}\lesssim
k_{B}T$, $E_{f}\lesssim k_{B}T$ according to Schatz $et~al.$,
including the ground state. $k_{B}$ is the Boltzmann constant, $T$
is the temperature in units of $\rm{K}$. We set $Q_{if}$ as the
energy difference between nuclei $(Z,A)$ and $(Z-1,A)$; thus,
$Q_{if}$ is always negative while $Q_{fi}$ is positive here,
$Q_{if}$ and $Q_{fi}$ in units of $\rm{MeV}$. This definition is
slightly different from the energy difference between parent nuclei
and daughter nuclei, but we think this helps avoid misunderstanding
when simultaneously dealing with the EC and $\beta^{-}$ decay
processes. Besides, we set $M_{(Z,A)}$ and $M_{(Z-1,A)}$ to be the
rest mass of the nuclei $(Z,A)$ and $(Z-1,A)$. We also define $W$ as
the total energy of the relativistic electrons in units of electron
rest mass $m_{e}c^2$. The threshold energy in units of electron rest
mass can be expressed as
\begin{equation}
q_{fi}=\frac{Q_{fi}}{m_{e}c^2},
Q_{fi}=M_{(Z-1,A)}c^2+E_{f}-M_{(Z,A)}c^2-E_{i}, q_{if}=-q_{fi}.
\end{equation}
For a degenerate relativistic Fermi electron gas, the rate formalism
as derived by Fuller, Fowler and Newman ( FFN )~\cite{Fuller:1985}
applies. Thus the electron-capture rate (per nucleus per unit time)
from state $i$ to state $f$ is
\begin{eqnarray}
\lambda^{+}& = &\frac{\rm ln2}{ft}\Phi^{+}(q_{if}),
\nonumber\\
\Phi^{+}(q_{if})& =
&\int_{w_{l}}^{\infty}F^+(Z,W)WP(W+q_{if})^2S(W)dW.
\end{eqnarray}
For the $\beta^{-}$ decay rate (per nucleus per unit time) from
state $f$ to state $i$, we have
\begin{eqnarray}
\lambda^-& = &\frac{\rm ln2}{ft}\Phi^-(q_{fi}),
\nonumber\\
\Phi^-(q_{fi})& = &\int_{1}^{w_{l}}F^-(Z,W)WP(q_{fi}-W)^2[1-S(W)]dW.
\end{eqnarray}
Similarly, the neutrino emissivity (energy per nucleus per unit time
emitted via neutrinos) from state $i$ to state $f$ is
\begin{eqnarray}
\xi^+& = &\frac{{\rm ln2}}{ft}m_{e}c^{2}\Psi^+(q_{if}),
\nonumber\\
\Psi^+(q_{if})& =
&\int_{w_{l}}^{\infty}F^{+}(Z,W)WP(W+q_{if})^3S(W)dW.
\end{eqnarray}
The neutrino emissivity (energy per nucleus per unit time emitted
via antineutrinos) from state $f$ to state $i$ is
\begin{eqnarray}
\xi^-& = &\frac{{\rm ln2}}{ft}m_{e}c^{2}\Psi^-(q_{fi}),
\nonumber\\
\Psi^-(q_{fi})& =
&\int_{1}^{w_{l}}F^{-}(Z,W)WP(q_{fi}-W)^3[1-S(W)]dW,
\end{eqnarray}
where $ft$ values measure the transition strength of EC and
$\beta^{-}$ decay processes, $F^\pm(Z,W)$ has been defined as the
Coulomb correction factor, $w_{l}=1$ if $q_{if}>-1$ or
$w_{l}=|q_{if}|$ if $q_{if}<-1$, and $S(W)$ is the Fermi-Dirac
distribution function. Since electrons are relativistic and
degenerate in accreting neutron stars crusts (except the outmost
part), their large energies make $F^{\pm}(Z,W)$ almost
constant~\cite{Beaudet:1972}. We take $\langle F\rangle^\pm\simeq
2\pi\alpha Z/|1-e^{\mp2\pi\alpha Z}|$, the average value of the
factor $F^\pm(Z,W)$ in our calculations, $\alpha$ is the fine
structure constant. $\Phi^{\pm}$ and $\Psi^{\pm}$ can be expressed
in what follows.

\begin{align}
&\Phi^{+}=(\frac{k_{B}T}{m_{e}c^2})^5 \langle
F\rangle^+[F_{4}(\eta)+2\chi F_{3}(\eta)+\chi^2F_{2}(\eta)],
\\
&\Phi^-=(\frac{k_{B}T}{m_{e}c^2})^5 \langle
F\rangle^-[F_{4}(-\eta)-2\chi F_{3}(-\eta)+\chi^2F_{2}(-\eta)],
\\
&\Psi^+=(\frac{k_{B}T}{m_{e}c^2})^6 \langle
F\rangle^+[F_{5}(\eta)+2\chi F_{4}(\eta)+\chi^2F_{3}(\eta)],
\\
&\Psi^-=(\frac{k_{B}T}{m_{e}c^2})^6 \langle
F\rangle^-[F_{5}(-\eta)-2\chi F_{4}(-\eta)+\chi^2F_{3}(-\eta)],
\end{align}
where $\chi=Q_{fi}/{(k_{B}T)}$, and
$\eta=(\mu_{e}+Q_{if})/(k_{B}T)=\delta\mu/(k_{B}T)$, keeping in mind
that $\chi$ is positive. The function $F_{k}(\eta)$ is a
relativistic Fermi integral of order $k$ defined as
\begin{equation}
F_{k}(\eta)=\int_{0}^{\infty}\frac{y^k}{1+e^{y-\eta}}dy.
\end{equation}

\section{ACCRETION-DRIVEN COMPRESSION ON URCA SHELL}
Schatz $et~al.$~\cite{Schatz:2014} proposed that the Urca shell
spans a range of electron chemical potential
$|Q_{if}|-k_{B}T\lesssim \mu_{e}\lesssim |Q_{if}|+k_{B}T$, the
thickness of the Urca shell defined by the thermal
fluctuations~\cite{Cooper:2009,Schatz:2014} is $(\Delta
R)_{\rm{shell}}\approx
|dR/dP_{e}|(dP_{e}/d\mu_{e})\Delta\mu_{e}\approx
(dP_{e}/d\mu_{e})(\Delta\mu_{e}/\rho g)$, where $R$ is the radius,
$P_{e}$ is the degenerate pressure of electrons, $\rho$ is the mass
density, and $g$ is the surface gravity. In this case,
$\Delta\mu_{e}=k_{B}T$, $(\Delta R)_{\rm{shell}}\approx
Y_{e}k_{B}T/(m_{u}g)$, where $Y_{e}\approx Z/A$ is the electron
fraction and $m_{u}$ is the atomic mass unit. EC and $\beta^{-}$
decay proceed equally in the Urca shell, taking away a large amount
of energy via neutrinos (and antineutrinos). For simplicity, we
assume the Urca shell consists of only nuclei $(Z,A)$ and $(Z-1,A)$,
with $n^{+}$ being the number density of $(Z,A)$ and $n^{-}$ being
the number density of $(Z-1,A)$. In chemical equilibrium state, the
steady state~\cite{Tsuruta:1970} gives
\begin{equation}
n^{+}\lambda^{+}=n^{-}\lambda^{-}.
\end{equation}
Combining Eqs. (2), (3), (6) and (7), it reads
\begin{eqnarray}
n^{+}\langle F \rangle^{+}[\chi^{2}F_{2}(0)+2\chi F_{3}(0)+F_{4}(0)]
\nonumber\\
=n^{-}\langle F \rangle^{-}[\chi^{2}F_{2}(0)-2\chi
F_{3}(0)+F_{4}(0)].
\end{eqnarray}
We define
\begin{equation}
\kappa(\chi)=\frac{n^{+}\langle F \rangle^{+}}{n^{-}\langle F
\rangle^{-}}=\frac{\chi^{2}F_{2}(0)-2\chi
F_{3}(0)+F_{4}(0)}{\chi^{2}F_{2}(0)+2\chi F_{3}(0)+F_{4}(0)}.
\end{equation}
For large $\chi$ value, which means the energy threshold is large or
the temperature is low, the $\chi^{2}F_{2}(0)$ term dominates, then
$\kappa(\chi)\approx 1$. TC70 has used this approximation in white
dwarfs with a temperature as low as $10^{4}\sim10^{5}~\rm{K}$. But
it is not appropriate for high temperature accreting neutron stars,
for example, the Urca pair
$^{81}_{35}\mathrm{Br}\,\textrm{--}\,^{81}_{34}\mathrm{Se}$ has a
ground-state to ground-state threshold energy
$|Q_{if}|=1.59~\rm{MeV}$, at the temperature
$T=4.5\times10^{8}~\rm{K}$, $\chi=41$, and $\kappa(\chi)=0.735$.
Therefore, the parameter $\kappa(\chi)$ should be considered.
Besides, the charge neutrality condition gives
\begin{equation}
n_{e}=Zn^{+}+(Z-1)n^{-}\simeq Z(n^{+}+n^{-}),
\end{equation}
where $n_{e}$ is the electron number density. Combining Eqs. (11)
and (14) one obtains
\begin{equation}
n^{+}\approx\frac{n_{e}\langle F \rangle^{-}\kappa(\chi)}{Z[\langle
F \rangle^{+}+\langle F \rangle^{-}\kappa(\chi)]},
~n^{-}\approx\frac{n_{e}\langle F \rangle^{+}}{Z[\langle F
\rangle^{+}+\langle F \rangle^{-}\kappa(\chi)]}.
\end{equation}
The net reaction rate is
\begin{eqnarray}
\Gamma(T,\eta) & = &n^{+}\lambda^{+}-n^{-}\lambda^{-}
\nonumber\\
&=&\frac{C}{k_{B}T}\Bigg[\chi^2\Bigg(F_{2}(\eta)-\frac{F_{2}(-\eta)}{\kappa(\chi)}\Bigg)
\nonumber\\
&&+2\chi\Bigg(F_{3}(\eta)+
\frac{F_{3}(-\eta)}{\kappa(\chi)}\Bigg)+\Bigg(F_{4}(\eta)-\frac{F_{4}(-\eta)}{\kappa(\chi)}\Bigg)\Bigg]
\nonumber\\
& \approx &
\frac{C}{k_{B}T}\chi^2\Bigg[F_{2}(\eta)-\frac{F_{2}(-\eta)}{\kappa(\chi)}\Bigg].
\end{eqnarray}
In the statistical equilibrium state, $\eta=0$ and $\Gamma(T,0)=0$,
the composition remains unchanged. The total neutrino energy-loss
rate is
\begin{eqnarray}
\epsilon_{\nu}(T,\eta)& = &n^+\xi^{+}+n^-\xi^{-}
\nonumber\\
&=&C\Bigg[\chi^2\Bigg(F_{3}(\eta)+\frac{F_{3}(-\eta)}{\kappa(\chi)}\Bigg)
\nonumber\\
&& +2\chi
\Bigg(F_{4}(\eta)-\frac{F_{4}(-\eta)}{\kappa(\chi)}\Bigg)+\Bigg(F_{5}(\eta)+\frac{F_{5}(-\eta)}{\kappa(\chi)}\Bigg)\Bigg]
\nonumber\\
& \approx
&C\chi^2\Bigg[F_{3}(\eta)+\frac{F_{3}(-\eta)}{\kappa(\chi)}\Bigg],
\end{eqnarray}
where
\begin{equation*}
C=\frac{{\rm ln2}}{ft}\frac{(k_{B}T)^6}{(m_{e}c^2)^5}
\frac{n_{e}\langle F\rangle^+\langle F
\rangle^-\kappa(\chi)}{Z~[\langle F \rangle^++\langle F
\rangle^-\kappa(\chi)]}.
\end{equation*}

The accretion-driven compression makes the Urca-shell scenario
different. We consider only the effect of gravitational compression
on the Urca shell under the weight of the newly accreted matter. On
the one hand, compression makes the Urca shell denser and denser,
resulting in an increase in electron number density. On the other
hand, the Urca shell may deviate from chemical equilibrium as a
response, resulting in a nonzero net reaction which tries to pull
the electron number density back to the original level. Therefore,
the Urca shell may be out of chemical equilibrium: we call this the
deviated shell. The compression and the net reaction rate compete to
determine the actual value $\delta\mu$, just like the case of
gravitational contraction of $npe$ (neutrons, protons and electrons)
matter that Reisenegger~\cite{Reisenegger:1995} discussed in
spin-down neutron stars. Or, in other words, the effect is the
consequence of the competition between the timescale $\tau_{com}$ to
increase $n_{e}$ by $\Delta n_{e}$ and the timescale $\tau_{weak}$
to consume electrons by $\Delta n_{e}$. The timescales $\tau_{com}$
and $\tau_{weak}$ are estimated in the following discussions.

We consider continual-accreting neutron stars, especially neutron
stars accreting at high rates. For example, the superbursting
sources accreting at $\dot{m}=(0.1-0.3)\dot{m}_{Edd}$. The weight of
newly accreted matter pushes the matter beneath it to higher density
and pressure; as a result, $n_{e}$ and $\mu_{e}$ keep rising. We can
make a rough estimate of the increasing rate of $n_{e}$:
$n_{e}\approx 4.3\times10^{30}\mu_{e}^{3}/\rm{cm^{3}}$, the pressure
is dominated by relativistic degenerate electrons in the crust and
has $P_{e}\approx
1.76\times10^{24}\rm{\mu_{e}^{4}}~\rm{erg/cm^{3}}$, $\mu_{e}$ in
units of $\rm{MeV}$, thus
\begin{equation}
\frac{dn_{e}}{dt}\approx\frac{dn_{e}}{dP_{e}}\frac{dP_{e}}{dt}=\frac{dn_{e}}{dP_{e}}\dot{m}g,
\end{equation}
where $g=GM/R^{2}$, and $M$ is the gravitational mass of the neutron
star. For the representative values $R=10~\rm{km}$ and
$M=1.4~M_{\odot}$, where $M_{\odot}$ is the mass of the sun,
$g=1.85\times10^{14}~\rm{cm}/\rm{s}^{2}$. We use
$g=1.85\times10^{14}~\rm{cm}/\rm{s}^{2}$ for the Urca pairs
throughout the crusts. Therefore,
\begin{equation}
\frac{dn_{e}}{dt}\approx
(3.43\times10^{25})\frac{\dot{m}/\dot{m}_{Edd}}{\mu_{e}}(\rm{cm^{-3}}~\rm{s^{-1}}).
\end{equation}
Equation (19) shows that $dn_{e}/dt$ is proportional to the local
accretion rate; thus, for superbursting sources, the high local
accretion rates will increase $n_{e}$ dramatically. The timescale to
increase electrons by $\Delta n_{e}$ is
\begin{equation}
\tau_{com}=\frac{\Delta n_{e}}{dn_{e}/dt}\propto
\frac{|Q_{if}|}{(3.43\times
10^{25})(\dot{m}/\dot{m}_{Edd})}~(\rm{s}),
\end{equation}
while the timescale to consume electrons by $\Delta n_{e}$ is
\begin{eqnarray}
\tau_{weak}(\eta)& =& \frac{\Delta n_{e}}{\Gamma(T,\eta)}\propto
\frac{Zft}{(5.42\times 10^{28})|Q_{if}|^{5}T_{9}^{3}\langle F
\rangle^{*}}
\nonumber\\
&& \times
\frac{1}{F_{2}(\eta)-F_{2}(-\eta)/\kappa(\chi)}\textbf{~(\rm{s}),}
\end{eqnarray}
where $\langle F \rangle^{*}=\langle F \rangle^{+}\langle F
\rangle^{-}\kappa(\chi)/[\langle F \rangle^{+}+\langle F
\rangle^{-}\kappa(\chi)]$, and $T_{9}=T/(10^9~\rm{K})$. As long as
the timescale satisfies the relation
\begin{equation}
\tau_{com}<\tau_{weak}(\eta),
\end{equation}
the Urca shell must be a deviated one, we then obtain the upper
limit of $\eta$ which identifies chemical departures from
$\tau_{com}=\tau_{weak}(\eta)$, thus
\begin{equation}
F_{2}(\eta)-\frac{F_{2}(-\eta)}{\kappa(\chi)}=(6.33\times10^{-4})\frac{Zft(\dot{m}/\dot{m}_{Edd})}{|Q_{if}|^{6}T_{9}^{3}\langle
F\rangle^{*}}.
\end{equation}

We have used the Fermi integrals we derived in the appendix.
Equation (23) shows that $\eta$ is more sensitive to temperature and
energy thresholds than $ft$ values. Besides, experimentally measured
$ft$ values have uncertainties. Hence, we fix $ft$ as a constant.
The existence of a real solution to Eq. (23) determines whether a
deviated shell is true. To make this specific, we discuss the effect
of compression on Urca pairs in Table
\uppercase\expandafter{\romannumeral1} of Deibel
$et~al.$~\cite{Deibel:2016}, which are identified as the strongest.
The majority of the corresponding $ft$ values fall in the range
$5.2<{\rm log}ft<5.7$. For example, $^{49}_{22}\rm Ti-^{49}_{21}\rm
Sc$ Urca pair has ${\log}ft=5.7$, $^{55}_{25}\rm Mn-^{55}_{24}\rm
Cr$ has ${\log}ft=5.2$ and $^{23}_{11}\rm Na-^{23}_{10}\rm Ne$ has
${\log}ft=5.3$~\cite{Tsuruta:1970}. For comparison, we take $\log
ft=5.3$ and $T_{9}=0.45$ in the following calculations.

\begin{table}[!h] \tabcolsep 1.9mm \caption{Ocean Urca pairs}
\begin{center}
\vspace{-0.1cm}
\renewcommand{\arraystretch}{1.2}
\begin{tabular}{r@{}l r@{.}l  r@{.}l  r@{.}l  r@{.}l}
\hline \hline \multicolumn{2}{c}{Urca Pair,$^A_{Z}\rm X$}
&\multicolumn{2}{c}{$|Q_{if}| [\rm MeV]$}
&\multicolumn{2}{c}{$\eta^{a} [k_{B}T]$}
&\multicolumn{2}{c}{$\eta^{b}[k_{B}T]$}
&\multicolumn{2}{c}{$\chi$}\\
\hline
&$^{81}_{35}\mathrm{Br}\,\textrm{--}\,^{81}_{34}\mathrm{Se}$          &1&59     &14&95   &18&94   &40&7      \\
&$^{49}_{22}\mathrm{Ti}\,\textrm{--}\,^{49}_{21}\mathrm{Sc}$          &2&00     &6&96    &8&98    &51&2      \\
&$^{65}_{29}\mathrm{Cu}\,\textrm{--}\,^{65}_{28}\mathrm{Ni}$          &2&14     &6&92    &8&93    &54&8      \\
&$^{55}_{25}\mathrm{Mn}\,\textrm{--}\,^{55}_{24}\mathrm{Cr}$          &2&60     &3&87    &5&21    &66&6      \\
&$^{69}_{30}\mathrm{Zn}\,\textrm{--}\,^{69}_{29}\mathrm{Cu}$          &2&68     &4&03    &5&40    &68&6      \\
&$^{57}_{26}\mathrm{Fe}^{\ast}\,\textrm{--}\,^{57}_{25}\mathrm{Mn}$   &2&70     &3&57    &4&84    &69&1      \\
&$^{67}_{29}\mathrm{Cu}\,\textrm{--}\,^{67}_{28}\mathrm{Ni}$          &3&58     &1&43    &2&62    &91&7      \\
&$^{63}_{28}\mathrm{Ni}^{\ast}\,\textrm{--}\,^{63}_{27}\mathrm{Co}$   &3&66     &1&25    &2&02    &93&7      \\
&$^{25}_{12}\mathrm{Mg}\,\textrm{--}\,^{25}_{11}\mathrm{Na}$          &3&83     &0&41    &0&73    &98&1      \\
&$^{81}_{34}\mathrm{Se}\,\textrm{--}\,^{81}_{33}\mathrm{As}$          &3&86     &1&24    &2&00    &98&8      \\
&$^{73}_{31}\mathrm{Ga}\,\textrm{--}\,^{73}_{30}\mathrm{Zn}$          &4&11     &0&80    &1&39    &105&2     \\
&$^{79}_{33}\mathrm{As}\,\textrm{--}\,^{79}_{32}\mathrm{Ge}$          &4&11     &0&87    &1&50    &105&2     \\
&$^{23}_{11}\mathrm{Na}\,\textrm{--}\,^{23}_{10}\mathrm{Ne}$          &4&38     &0&20    &0&34    &112&1     \\
&$^{101}_{42}\mathrm{Mo}^{\ast}\,\textrm{--}\,^{101}_{41}\mathrm{Nb}$ &4&63     &0&69    &1&21    &118&5     \\
&$^{57}_{25}\mathrm{Mn}\,\textrm{--}\,^{57}_{24}\mathrm{Cr}$          &4&96     &0&23    &0&41    &127&0     \\
\hline
\end{tabular}
\footnotetext[1]{Calculated with $\log ft=5.3$, $T_{9}=0.45$ and
$\dot{m}=0.1~\dot{m}_{Edd}$.} \footnotetext[2]{Calculated with $\log
ft=5.3$, $T_{9}=0.45$ and $\dot{m}=0.2~\dot{m}_{Edd}$.}
\end{center}
\end{table}

\begin{figure}
\centering
\resizebox{\hsize}{!}{\includegraphics[width=\linewidth]{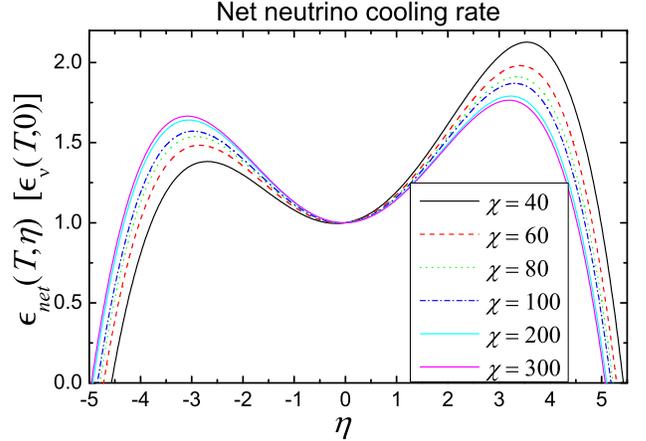}}
         \caption{ Net cooling rate $\epsilon_{net}(T,\eta)$ in units of $\epsilon_{\nu}(T,0)$. According to the $\chi$ values in Table \uppercase\expandafter{\romannumeral1},
         we choose $\chi=40,60,80,100,200,300$ for illustrations.
         These lines have shown much in common, the dependence of net cooling rate on $\chi$ values (thus Urca pairs) is relatively weak, the net cooling rate will
         be enhanced if $0<\eta\lesssim4.5$, when $\eta>5.5$, there will be no net cooling but net heating.}
   \label{fig:1}
\end{figure}

Table \uppercase\expandafter{\romannumeral1} shows the upper limits
of possible chemical departure values at the local accretion rate
$\dot{m}=0.1\dot{m}_{Edd}$ and $\dot{m}=0.2\dot{m}_{Edd}$. In
chemical departure cases, the net chemical energy release rate is
$\Gamma(T,\eta)\delta\mu$, and the total neutrino energy-loss rate
is $\epsilon_{\nu}(T,\eta)=n^+\xi^{+}+n^-\xi^{-}$, thus the net
cooling rate is
$\epsilon_{net}(T,\eta)=\epsilon_{\nu}(T,\eta)-\Gamma(T,\eta)\delta\mu$.
The total cooling rate in the shell~\cite{Deibel:2016} is
\begin{equation}
L(\eta)=4\pi R^{2}\int_{\rm{shell}}\epsilon_{net}(T,\eta)dz',
\end{equation}
where $dz'=(dP_{e}/d\mu_{e})[d\mu_{e}/(\rho g)]$. We assume the
deviated shell has the same composition and thickness with the Urca
shell in chemical equilibrium. It is hard to tell the relation
between the net chemical energy release rate
$\Gamma(T,\eta)\delta\mu$ and the total neutrino energy-loss rate
$\epsilon_{\nu}(T,\eta)$ according to Eqs. (16) and (17). When the
approximation $\kappa(\chi)\approx 1$ is taken, Eqs. (16) and (17)
can be expressed in units of $\epsilon_{\nu}(T,0)$ analytically,
that
\begin{equation}
\Gamma(T,\eta)\delta\mu=\Gamma(T,\eta)\times
k_{B}T\eta\simeq\epsilon_{\nu}(T,0)(\frac{20 \eta ^2}{7 \pi
^2}+\frac{20 \eta ^4}{7 \pi ^4}),
\end{equation}
and
\begin{equation}
\epsilon_{\nu}(T,\eta)\simeq\epsilon_{\nu}(T,0)(1+\frac{30 \eta
^2}{7 \pi ^2}+\frac{15 \eta ^4}{7 \pi ^4}),
\end{equation}
where $\epsilon_{\nu}(T,0)$ is the total neutrino energy-loss rate
in chemical equilibrium. It is clear that the net chemical energy
release rate increases faster than the total neutrino energy-loss
rate, the heating and cooling compete, to a certain extent, net
cooling will be replaced by net heating. The net cooling rate is
presented in Fig. 1, which is plotted numerically according to Eqs.
(16) and (17).

\section{Results AND CONCLUSIONS}
We have investigated the effect of compression on Urca shells in
fast accreting neutron stars. We find that, in neutron stars with
local accretion rates $\dot{m}=(0.1-0.3)\dot{m}_{Edd}$, if the Urca
shell exists in the ocean and crust of neutron stars, compression
may make the Urca shell deviate from chemical equilibrium. The upper
limits of chemical departure values for 15 strongest Urca pairs,
which are calculated based on reasonable parameters in superbursting
sources, are presented in Table
\uppercase\expandafter{\romannumeral1}. Generally, the shallower the
Urca shell exists, the bigger the upper limit of the chemical
departure value is. At $\dot{m}=(0.1-0.3)\dot{m}_{Edd}$, effect of
compression on Urca pairs with $|Q_{if}|>4~\rm{MeV}$ will be very
small, this means the Urca shell in the deeper crust (with higher
$|Q_{if}|\approx 10~\rm{MeV}$) may always be in chemical equilibrium
even at the local Eddington accretion rate case.

As stated in the previous section, we assume the deviated shell has
the same composition as that in chemical equilibrium. Under this
assumption, we numerically calculated the net cooling rate. Figure 1
shows that the net cooling rate will be enhanced when
$0<\eta\lesssim4.5$, thus chemical departures may also make
contributions to the total neutrino (and antineutrino) luminosity
besides the Urca shell in chemical equilibrium. For the temperature
(about $4\times10^{8}~\rm{K}$), the local accretion rate
[$\dot{m}=(0.1-0.3)\dot{m}_{Edd}$] and Urca pairs (in Table
\uppercase\expandafter{\romannumeral1}) we are interested in,
$\epsilon_{net}(T,\eta)$ will be enhanced by a factor of $1\sim 2$.
What is more, there will be no net cooling but net heating when
$\eta\gtrsim 5.5$. Table \uppercase\expandafter{\romannumeral1}
shows that the upper limits of
$^{81}_{35}\mathrm{Br}\,\textrm{--}\,^{81}_{34}\mathrm{Se}$,
$^{49}_{22}\mathrm{Ti}\,\textrm{--}\,^{49}_{21}\mathrm{Sc}$ and
$^{65}_{29}\mathrm{Cu}\,\textrm{--}\,^{65}_{28}\mathrm{Ni}$ are
slightly bigger than $5.5$, which indicates that Urca pairs in
shallow depth tend to result in net heating rather than cooling.
Although the effect of heating can be pretty small, it is
significant for accreting neutron stars because the crust Urca shell
in chemical equilibrium proposed by Schatz $et~al.$ always acts as a
parameterized cooling source. However, in consideration of the
difference in accretion rate and threshold energy of Urca pairs,
formation of a deviated shell makes the Urca-shell scenario differ
from star to star and from Urca pair to Urca pair.

We conclude that, in accreting neutron stars, especially neutron
stars accreting at a few tens of the local Eddington accretion rate,
the Urca shell may deviate from chemical equilibrium, to what degree
the Urca shell is deviated is determined by the properties of the
nuclei, the temperature, and the actual local accretion rate. It is
easier for Urca pairs in shallow depth to be deviated. The net
cooling rate of these Urca pairs would be slightly enhanced if
$0<\eta\lesssim4.5$; however, net heating will appear if
$\eta\gtrsim 5.5 $. Only the strongest Urca pairs at deeper depth
may always be in chemical equilibrium, indicating that these Urca
pairs are most stable.

Deibel $et~al.$ have not considered Urca pairs with large $\log ft$
because neutrino emissivity of these pairs are much too small. We
think, however, that Urca pairs with large $\log ft$ have more
chances to be deviated from chemical equilibrium because the upper
limits of $\eta$ can be larger. Does this mean that Urca shells are
not stable structures in accreting neutron stars? We do not know
yet; more constraints should be put on the actual $\eta$ (or
$\delta\mu$). We hope this work may help to further understand some
relevant astrophysical phenomena, such as
Type-\uppercase\expandafter{\romannumeral1} x-ray bursts and
superbursts. For more careful work, we think the effect of diffusion
and convection~\cite{Medin:2011,Medin:2014,Medin:2015} should also
be discussed, but accretion-driven compression is certainly
important.

\section*{\uppercase {acknowledgments}}
The authors would like to thank the anonymous referee very much for
helpful comments, which have greatly improved our work. This work is
supported by the National Natural Science Foundation of China (Grant
No. 11178001) and CCNU-QLPL Innovation Fund (Grant No. QLPL2015P01).

\section*{\uppercase {Appendix}}
The Fermi integrals may be calculated many times in the stellar
evolution codes~\cite{Fuller:1985}, thus simple and analytical
expressions for them are needed. FFN had obtained good
approximations for them by using the differential recursion relation
$dF_{k}(\eta)/d\eta=kF_{k-1}(\eta)$ and $F_{0}(\eta)=\rm
ln(1+e^{\eta})$. Based on these, they give
\begin{align}
&F_{1}(\eta)+F_{1}(-\eta)=\frac{\eta^2}{2}+\frac{\pi^2}{6},
\\
&F_{2}(\eta)-F_{2}(-\eta)=\frac{\eta^3}{3}+\frac{\pi^2\eta}{3},
\\
&F_{3}(\eta)+F_{3}(-\eta)=\frac{\eta^4}{4}+\frac{\pi^2\eta^2}{2}+\frac{7\pi^4}{60},
\\
&F_{4}(\eta)-F_{4}(-\eta)=\frac{\eta^5}{5}+\frac{2\pi^2\eta^3}{3}+\frac{7\pi^4\eta}{15},
\\
&F_{5}(\eta)+F_{5}(-\eta)=\frac{\eta^6}{6}+\frac{5\pi^2\eta^4}{6}+\frac{7\pi^4\eta^2}{6}+\frac{31\pi^6}{126},
\end{align}

and
\begin{align}
&F_{1}(\eta)=\left\{
\begin{array}{ll}
e^{\eta},~~~~~~~~~~~~~~~~~~\eta\leqslant0,\\
\frac{\eta^2}{2}+2-e^{-\eta},~~\eta>0,\\
\end{array}\right.\\
&F_{2}(\eta)=\left\{
\begin{array}{ll}
2e^{\eta},~~~~~~~~~~~~~~~~~~~~~~\eta\leqslant0,\\
\frac{\eta^3}{3}+\frac{\pi^{2}\eta}{3}+2e^{-\eta},~~~\eta>0,\\
\end{array}\right.\\
&F_{3}(\eta)=\left\{
\begin{array}{ll}
6e^{\eta},~~~~~~~~~~~~~~~~~~~~~~~~~~~~~~\eta\leqslant0,\\
\frac{\eta^4}{4}+\frac{\pi^2\eta^2}{2}+12-6e^{-\eta},~~\eta>0,\\
\end{array}\right.\\
&F_{4}(\eta)=\left\{
\begin{array}{ll}
24e^{\eta},~~~~~~~~~~~~~~~~~~~~~~~~~~~~~~~~~~\eta\leqslant0,\\
\frac{\eta^5}{5}+\frac{2\pi^2\eta^3}{3}+48\eta+24e^{-\eta},~~\eta>0,\\
\end{array}\right.\\
&F_{5}(\eta)=\left\{
\begin{array}{ll}
120e^{\eta},~~~~~~~~~~~~~~~~~~~~~~~~~~~~~~~~~~~~~~~~~~~~~~\eta\leqslant0,\\
\frac{\eta^6}{6}+\frac{5\pi^2\eta^4}{6}+\frac{7\pi^4\eta^2}{6}+240-120e^{-\eta},~~\eta>0.\\
\end{array}\right.
\end{align}
Their results are algebraically simple when $|\eta|\gg 0$, however,
the largest error for $F_{k}(\eta)(\eta=1,2,3,4,5)$ in the range
$|\eta| \leq 5$ can be up to $20\%$, especially when $\eta\approx
0$. Here we propose an improvement of these Fermi integrals.

We do it by changing the integration limits. For odd $k$,
\begin{equation}
F_{k}(\eta)+F_{k}(-\eta)=
\int_{0}^{\infty}\frac{(y+\eta)^{k}+(y-\eta)^{k}}{1+e^y}dy-\int_{0}^{\eta}(y-\eta)^{k}dy,
\end{equation}
and for even $k$,

\begin{equation}
F_{k}(\eta)-F_{k}(-\eta)=
\int_{0}^{\infty}\frac{(y+\eta)^{k}-(y-\eta)^{k}}{1+e^y}dy+\int_{0}^{\eta}(y-\eta)^{k}dy.
\end{equation}
According to Eqs. (37) and (38), we can reproduce the results in
Eqs.(27)-(31). Besides, for odd $k$,
\begin{eqnarray}
F_{k}(\eta)-F_{k}(-\eta) & = &
\int_{0}^{\infty}\frac{(y+\eta)^{k}-(y-\eta)^{k}}{1+e^y}dy
\nonumber\\
&&-\int_{0}^{\eta}\frac{(e^y-1)(y-\eta)^{k}}{1+e^y}dy,
\end{eqnarray}
and for even $k$,
\begin{eqnarray}
F_{k}(\eta)+F_{k}(-\eta)& = &
\int_{0}^{\infty}\frac{(y+\eta)^{k}-(y-\eta)^{k}}{1+e^y}dy
\nonumber\\
&&+\int_{0}^{\eta}\frac{(e^y-1)(y-\eta)^{k}}{1+e^y}dy.
\end{eqnarray}
When $\mid \eta \mid<2$, the factor $(e^y-1)/(e^y+1)$ can be
expressed by Taylor expansion around $\eta=0$,
\begin{equation}
\frac{e^y-1}{e^y+1}\approx
\frac{y}{2}-\frac{y^3}{24}+\frac{y^5}{240}.
\end{equation}
Keeping the first two terms in Eq. (41) and Combining Eqs.
(27)-(31), we arrive at
\begin{eqnarray}
F_{1}(\eta)& =
&\frac{\pi^2}{12}+(\ln{2})\eta+\frac{\eta^2}{4}+\frac{\eta^3}{24}-
\frac{\eta^{5}}{960},
\nonumber\\
F_{2}(\eta)& =
&\frac{3}{2}\zeta(3)+\frac{\pi^2\eta}{6}+(\ln{2})\eta^2+
\frac{\eta^3}{6}+\frac{\eta^4}{48}-\frac{\eta^{6}}{2880},
\nonumber\\
F_{3}(\eta)& = &
\frac{7\pi^4}{120}+\frac{9}{2}\zeta(3)\eta+\frac{\pi^2\eta^2}{4}+(\ln{2})\eta^3
+\frac{\eta^4}{8}+\frac{\eta^5}{80}-\frac{\eta^{7}}{6720},
\nonumber\\
F_{4}(\eta)& =
&\frac{45\zeta(5)}{2}+\frac{7\pi^4\eta}{30}+9\zeta(3)\eta^2+\frac{\pi^2\eta^2}{3}
+(\ln{2})\eta^4+\frac{\eta^5}{10}
\nonumber\\
&& +\frac{\eta^6}{120}-\frac{\eta^{8}}{13440},
\nonumber\\
F_{5}(\eta)& =
&\frac{31\pi^4}{256}+\frac{225}{2}\zeta(5)\eta+\frac{7\pi^4}{12}\eta^2+
15\zeta(3)\eta^3+\frac{5\pi^2}{12}\eta^4
\nonumber\\
&&
+(\ln2)\eta^5+\frac{\eta^6}{12}+\frac{\eta^7}{168}-\frac{\eta^{9}}{24192},
\end{eqnarray}
where $\zeta$ represent the Riemann zeta function,
$\zeta(3)=1.20206$ and $\zeta(5)=1.03693$.

Finally, we get the complete expressions of Fermi integrals of order
$k$ ($k=1-5$),
\begin{align}
&F_{1}(\eta)=\left\{
\begin{array}{ll}
e^{\eta},~~~~~~~~~~~~~~~~~~~~~~~~~~~~~~~~~~~~~~~~~~~~~~~~~~~~~~~~~\eta\leqslant-2,\\
\frac{\pi^2}{12}+(\ln{2})\eta+\frac{\eta^2}{4}+\frac{\eta^3}{24}-
\frac{\eta^{5}}{960},~~~~~~~~~~~~~~~~|\eta|<2,\\
\frac{\eta^2}{2}+2-e^{-\eta},~~~~~~~~~~~~~~~~~~~~~~~~~~~~~~~~~~~~~~~~~\eta\geq 2,\\
\end{array}\right.\\
&F_{2}(\eta)=\left\{
\begin{array}{ll}
2e^{\eta},~~~~~~~~~~~~~~~~~~~~~~~~~~~~~~~~~~~~~~~~~~~~~~~~~~~~~~~\eta\leqslant-2,\\
\frac{3}{2}\zeta(3)+\frac{\pi^2\eta}{6}+(\ln{2})\eta^2+
\frac{\eta^3}{6}+\frac{\eta^4}{48}
\\-\frac{\eta^{6}}{2880},~~~~~~~~~~~~~~~~~~~~~~~~~~~~~~~~~~~~~~~~~~~~~~~~~~~|\eta|<2,\\
\frac{\eta^3}{3}+\frac{\pi^{2}\eta}{3}+2e^{-\eta},~~~~~~~~~~~~~~~~~~~~~~~~~~~~~~~~~~~~\eta\geq 2,\\
\end{array}\right.\\
&F_{3}(\eta)=\left\{
\begin{array}{ll}
6e^{\eta},~~~~~~~~~~~~~~~~~~~~~~~~~~~~~~~~~~~~~~~~~~~~~~~~~~~~~~~\eta\leqslant-2,\\
\frac{7\pi^4}{120}+\frac{9}{2}\zeta(3)\eta+\frac{\pi^2\eta^2}{4}+(\ln{2})\eta^3
\\+\frac{\eta^4}{8}+\frac{\eta^5}{80}-\frac{\eta^{7}}{6720},~~~~~~~~~~~~~~~~~~~~~~~~~~~~~~~~~~~|\eta|<2, \\
\frac{\eta^4}{4}+\frac{\pi^2\eta^2}{2}+\frac{7\pi^{4}}{60}-6e^{-\eta},~~~~~~~~~~~~~~~~~~~~~~~~~~\eta\geq
2,\\
\end{array}\right.\\
&F_{4}(\eta)=\left\{
\begin{array}{ll}
24e^{\eta},~~~~~~~~~~~~~~~~~~~~~~~~~~~~~~~~~~~~~~~~~~~~~~~~~~~~~\eta\leqslant-2,\\
\frac{45\zeta(5)}{2}+\frac{7\pi^4\eta}{30}+9\zeta(3)\eta^2+\frac{\pi^2\eta^2}{3}
\\+(\ln{2})\eta^4+\frac{\eta^5}{10}
+\frac{\eta^6}{120}-\frac{\eta^{8}}{13440},~~~~~~~~~~~~~~~~|\eta|<2,\\
\frac{\eta^5}{5}+\frac{2\pi^2\eta^3}{3}+\frac{7\pi^{4}\eta}{15}+24e^{-\eta},~~~~~~~~~~~~~~~~~~~~~\eta\geq
2,\\
\end{array}\right.\\
&F_{5}(\eta)=\left\{
\begin{array}{ll}
120e^{\eta},~~~~~~~~~~~~~~~~~~~~~~~~~~~~~~~~~~~~~~~~~~~~~~~~~~~~\eta\leqslant-2,\\
\frac{31\pi^4}{256}+\frac{225}{2}\zeta(5)\eta+\frac{7\pi^4}{12}\eta^2+
15\zeta(3)\eta^3 \\+\frac{5\pi^2}{12}\eta^4
+(\ln2)\eta^5+\frac{\eta^6}{12}+\frac{\eta^7}{168}-\frac{\eta^{9}}{24192},~~~~|\eta|<2,\\
\frac{\eta^6}{6}+\frac{5\pi^2\eta^4}{6}+\frac{7\pi^4\eta^2}{6}+\frac{31\pi^{6}}{126}-120e^{-\eta}.~~~~~~~~~\eta\geq 2.\\
\end{array}\right.
\end{align}

The above Fermi integral results for $F_{1}(\eta)$ through
$F_{5}(\eta)$ are asymptotically exact for $|\eta|\geq 2$. The
largest error appears around $\eta=-2$, about $3.94\%$ for
$F_{1}(\eta\rightarrow-2^{-})$, $1.65\%$ for $F_{2}(\eta=-2)$,
$0.8\%$ for $F_{3}(\eta=-2)$, $0.4\%$ for $F_{4}(\eta=-2)$ and
$0.2\%$ for $F_{5}(\eta=-2)$. Thus we have got fine segmenting
functions for Fermi integrals $F_{k}(\eta)$ of order $k=1,2,3,4,5$.
It should be pointed out that these segmenting functions are not
continuous at $\eta=\pm2$. But when we are focusing on simplicity
and accuracy, this improvement may be helpful.

\end{spacing}

\begin{thebibliography}{}

\bibitem{Cornelisse:2000}
R. Cornelisse, J. Heise, E. kuulkers, F. Verbunt, and J. J. M. In't
Zand, Astron. Astrophys. 357, L21 (2000).


\bibitem{Zand:2004}
J. J. M. in't Zand, R. Cornelisse, and A. Cumming, Astron.
Astrophys. 426, 257 (2004).



\bibitem{Strohmayer:2002}
T. E. Strohmayer and E. F. Brown, Astrophys. J. 566, 1045 (2002).



\bibitem{Wijnands:2001}
R. Wijnands, Astrophys. J. 554, L59 (2001).



\bibitem{Haensel:1990}
P. Haensel and J. L. Zdunik, Astron. Astrophys. 227, 431 (1990).

\bibitem{Haensel:2003}
P. Haensel and J. L. Zdunik, Astron. Astrophys. 404, L33 (2003).

\bibitem{Gupta:2007}
S. Gupta, E. F. Brown, H. Schatz, P. M\"{o}ller, and K-L. Kratz,
Astrophys. J. 662, 1188 (2007).

\bibitem{Haensel:2008}
P. Haensel, J. L. Zdunik, Astron. Astrophys. 480, 459 (2008).




\bibitem{Woosley:1976}
S. E. Woosley and R. E. Taam, Nature (London) 263, 101 (1976).

\bibitem{Taam:1978}
R. E. Taam and R. E. Picklum, Astrophys. J. 224, 210 (1978).

\bibitem{Brown:1998}
E. F. Brown and L. Bildsten, Astrophys. J. 496, 915 (1998).

\bibitem{Cumming:2001}
A. Cumming and L. Bildsten, Astrophys. J. 559, L127 (2001).


\bibitem{Cooper:2009}
R. L. Cooper, A. W. Steiner, and E. F. Brown, Astrophys. J. 702, 660
(2009).



\bibitem{Cumming:2006}
A. Cumming, J. Macbeth, J. J. M. In't Zand, and D. Page, Astrophys.
J. 646, 429 (2006).




\bibitem{Caughlan:1988}
G. R. Caughlan and W. A. Fowler, At. Data Nucl. Data Tables 40, 283
(1988).




\bibitem{Keek:2012}
K. Keek, A. Heger, and J. J. M. In't Zand, Astrophys. J. 752, 150
(2012).




\bibitem{Brown:2004}
E. F. Brown, Astrophys. J. 614, L57 (2004).



\bibitem{Cooper:2005}
R. L. Cooper and R. Narayan, Astrophys. J. 629, 422 (2005).


\bibitem{Schatz:2014}
H. Schatz, S. Gupta, P. M\"{o}ller, M. Beard, E. F. Brown, A. T.
Deibel, L. R. Gasques, W. R. Hix, L. Keek, R. Lau, A. W. Steiner and
M. Wiescher, Nature (London) 505, 62 (2014).



\bibitem{Gamow:1941}
G. Gamow and M. Schoenberg, Phys. Rev. 59, 539 (1941).


\bibitem{Deibel:2016}
A. Deibel, Z. Meisel, H. Schatz, E. F. Brown and A. Cumming, arXiv,
1603. 01281v1.



\bibitem{Tsuruta:1970}
S. Tsuruta and A. G. W. Cameron, Astrophys. Space Sci. 7, 374
(1970).


\bibitem{Paczynski:1972}
B. Paczynski, Astrophys. Lett. 11, 53 (1972).



\bibitem{Woosley:1986}
S. E. Woosley and T. A. Weaver, Ann. Rev. Astro. Astrophys. 24, 205
(1986).



\bibitem{Keek:2011}
L. Keek and A. Heger, Astrophys. J. 743,189 (2011).



\bibitem{Woosley:2004}
S. E. Woosley, A. Heger, A. Cumming, R. D. Hoffman, J. Pruet, T.
Rauscher, J. L. Fisker, H. Schatz, B. A. Brown and M. Wiescher,
Astrophys. J. Suppl. Ser. 151, 75 (2004).




\bibitem{Fuller:1985}
G. M. Fuller, W. A. Fowler and M. J. Newman, Astrophys. J. 293, 1
(1985).


\bibitem{Beaudet:1972}
G. Beaudet, E. E. Salpeter, and M. L. Silvestro, Astrophys. J.
174,79 (1972).


\bibitem{Reisenegger:1995}
A. Reisenegger, Astrophys. J. 442, 749 (1995).



\bibitem{Medin:2011}
Z. Medin and A. Cumming, Astrophys. J. 730, 97 (2011).


\bibitem{Medin:2014}
Z. Medin and A. Cumming, Astrophys. J. Lett. 783, L3 (2014).



\bibitem{Medin:2015}
Z. Medin and A. Cumming, Astrophys. J. 702, 29 (2015).




\end{thebibliography}
\end{document}